\documentclass[final,5p,times,twocolumn,number]{elsarticle}  

\biboptions{square}  

\usepackage{amssymb}
\usepackage{lipsum}
\usepackage{comment}
\usepackage{url}
\usepackage{graphicx}
\usepackage{booktabs}
\usepackage{multirow}
\usepackage{amsmath}
\usepackage{tabularx} 
\usepackage{caption}  
\usepackage{adjustbox} 
\makeatletter
\renewcommand{\@cite}[2]{[{#1}]} 
\renewcommand\@biblabel[1]{[#1]} 
\makeatother

\begin{document}

\begin{frontmatter}



\title{Utilizing RNN for Real-time Cryptocurrency Price Prediction and Trading Strategy Optimization}

\author{Shamima Nasrin Tumpa\corref{cor1}}
\ead{stumpa42@tntech.edu}
\address{Department of Mathematics, Tennessee Tech University, Cookeville, TN, USA}

\author{Kehelwala Dewage Gayan Maduranga}
\ead{gmaduranga@tntech.edu}
\address{Department of Mathematics, Tennessee Tech University, Cookeville, TN, USA}

\cortext[cor1]{Corresponding author}

\begin{abstract}
This research embarks on the exploration of leveraging Recurrent Neural Network (RNN) for real-time cryptocurrency price prediction and the optimization of trading strategies. Despite the notorious volatility and unpredictability of the cryptocurrency market, traditional forecasting methods and trading strategies often fail to deliver desired results. This study aims to bridge this gap by harnessing the power of RNN, renowned for their proficiency in capturing long-term dependencies and trends in time-series data. 
Over a concentrated period of ten weeks, the project unfolds through a series of meticulously planned phases, starting with a comprehensive review of the existing literature and the collection of extensive datasets encompassing historical price data, trading volumes and sentiment analysis derived from social networks and news sources. The subsequent weeks are dedicated to data preprocessing, feature engineering, and the iterative development and refinement of the RNN model to accurately predict cryptocurrency prices. This foundation paves the way for the formulation of dynamic trading strategies that are rigorously backtested to assess profitability and risk, culminating in an evaluation phase in which the efficacy of the model and the performance of the trading strategies are thoroughly analyzed.
The anticipated outcome of this research is a robust RNN-based predictive model that not only surpasses traditional forecasting methods in accuracy but also empowers traders with optimized strategies tailored for the fast-paced cryptocurrency market. This study not only contributes to the academic discourse on financial market predictions using deep learning techniques, but also offers practical insights and tools for investors navigating the complexities of cryptocurrency trading. Through this endeavor, our goal is to set a precedent for future research to integrate advanced machine learning models with financial trading systems to navigate and profit from the digital currency ecosystem.

\end{abstract}

\begin{keyword}
Cryptocurrency, Long Short Term Memory (LSTM), Gated Recurrent Unit (GRU), Bidirectional Long Short Term Memory (Bi-LSTM), Root Mean Squared Error 
(RMSE) , Mean Absolute Percentage Error (MAPE).\\

This work was presented at the Actuarial Research Conference (ARC) 2024.Research abstract submitted and presented at ARC 2024. More details can be found at \url{https://sites.google.com/view/arc2024/home}.

\end{keyword}
\end{frontmatter}




\section{Introduction}
\label{introduction}
The foundation of modern financial systems is predominantly based on fiat currency, which has notable benefits such as divisibility, durability, and ease of transferability. However, fiat currency, being unbacked by a tangible asset, can lead to inflationary pressures and manipulation by centralized authorities, like governments. The centralized control of the monetary supply has historically caused issues like hyperinflation and rising income inequality \cite{Ahamad2022}. Additionally, the financial system's dependence on intermediaries like banks and credit card companies introduces higher costs, delays, and the risk of security breaches \cite{Gandal2018}. This reliance results in the loss of individuals' control over personal data.

Despite the limitations, trust in the current financial system is supported by government regulations and legal contracts. However, trust has been breached in the past due to events such as the dot-com bubble in the late 1990s and the 2008 real estate crisis, which led to significant financial losses \cite{Urquhart2016}. These events highlight the need for a more secure, transparent financial system. In 2008, an anonymous entity known as Satoshi Nakamoto introduced blockchain technology, along with the first decentralized cryptocurrency, Bitcoin (BTC), which enabled peer-to-peer (P2P) transactions without the need for third-party intermediaries \cite{Nakamoto2008}. Blockchain technology has since gained traction across various industries and has become a subject of extensive research \cite{Wang2020}.

Cryptocurrencies, powered by blockchain technology, have emerged as a new form of digital currency that employs cryptography to secure and verify transactions \cite{Ahamad2022}. Unlike fiat currencies, cryptocurrencies operate on decentralized networks without the oversight of a central authority. Bitcoin, the first cryptocurrency, is the most widely recognized, but other digital currencies, including Ethereum (ETH), Litecoin (LTC), and Ripple (XRP), have also gained prominence. Ethereum, in particular, introduced smart contracts and decentralized applications (dApps), broadening the use cases of blockchain technology \cite{Ahamad2022}.

A unique feature of cryptocurrencies is their extreme price volatility, which makes the market both lucrative and risky for investors. As the market matures, artificial intelligence (AI) and machine learning (ML) have been used to predict price movements. However, predicting cryptocurrency prices remains challenging due to the complexity of market drivers, such as government regulations, technological innovations, and public sentiment \cite{Gandal2018}. Despite these challenges, the cryptocurrency market is expected to continue growing, with forecasts estimating a compound annual growth rate (CAGR) of 11.1\% \cite{Ahamad2022}.

To improve prediction accuracy, this paper applies deep learning (DL) techniques to identify hidden patterns in cryptocurrency price data. By employing advanced DL algorithms, we aim to enhance prediction models, enabling more informed investment decisions. Specifically, the contributions of this paper are:
\begin{itemize}
    \item Developing a price prediction model for Bitcoin (BTC), Ethereum (ETH), and Litecoin (LTC);
    \item Utilizing DL algorithms such as Long Short-Term Memory (LSTM), Bi-directional LSTM (Bi-LSTM), and Gated Recurrent Units (GRU);
    \item Evaluating the models using performance metrics such as Mean Squared Error (MSE), Mean Absolute Error (MAE), Root Mean Squared Error (RMSE) and Mean Absolute Percentage Error (MAPE).
\end{itemize}

The objective is to create reliable models that cryptocurrency investors and traders can use for price predictions based on historical data.

\section{Literature Review}
\label{Literature Review}

Machine learning (ML) has emerged as a key area of artificial intelligence that involves predicting future events based on historical data. Specifically, in the domain of cryptocurrency price prediction, ML models have been applied to forecast market movements, showing enhanced accuracy over traditional financial prediction methods \cite{Ferdiansyah2023}. Numerous techniques like decision trees, support vector machines (SVM), and neural networks (NN) have proven effective for time-series forecasting, especially when predicting the price of digital currencies such as Bitcoin (BTC), Ethereum (ETH), and Litecoin (LTC).

Several studies have demonstrated the efficacy of ML algorithms in cryptocurrency prediction. For instance, one study compared multiple ML models like SVM, Artificial Neural Networks (ANN), and deep learning (DL) for predicting prices of cryptocurrencies including BTC and ETH, concluding that SVM was the most accurate approach \cite{Muniye2021}. In another study, Long Short-Term Memory (LSTM) models, which are a specialized form of recurrent neural networks (RNNs), were found to produce the lowest prediction error for BTC prices \cite{Muniye2021}.

Beyond individual models, ensemble methods have also been explored in recent research. One such study combined ANN, K-nearest neighbors (KNN), and gradient-boosted trees to predict the prices of nine cryptocurrencies, showing that ensemble models outperformed standalone models in minimizing prediction errors \cite{Seabe2022}. In a different study, an ensemble of random forests (RF) and Gradient Boosting Machine (GBM) was used to predict the prices of BTC, ETH, and Ripple (XRP), achieving mean absolute percentage error (MAPE) values between 0.92\% and 2.61\% \cite{Alimohammadi2021}.

In recent years, deep learning models have garnered significant attention due to their ability to process large datasets and discover hidden patterns. In particular, LSTM\cite{Hochreiter1997_LSTM} and Gated Recurrent Units (GRU)\cite{Cho2014_GRU}, which are types of RNNs, have proven particularly adept at financial time-series prediction \cite{Ferdiansyah2023}. A two-stage approach was proposed in one study where ANN and RF models were used to identify relevant features for BTC price prediction before employing LSTM for final forecasts. This method was shown to outperform classical models like ARIMA and SVM \cite{Ferdiansyah2023}.

Additionally, hybrid models have also been investigated. For instance, a model combining LSTM and GRU for predicting LTC and Monero (XMR) prices achieved higher accuracy compared to single-model approaches like LSTM \cite{Alimohammadi2021}. Furthermore, recent studies have proposed using convolutional neural networks (CNN) alongside LSTM to improve the accuracy of cryptocurrency price predictions. One study introduced a novel ensemble of LSTM, Bi-LSTM \cite{Schuster1997_BiLSTM}, and CNN for predicting hourly BTC, ETH, and XRP prices, achieving highly accurate results \cite{Seabe2022}.

These findings indicate the growing efficacy of deep learning models, especially when hybrid approaches are used to improve prediction accuracy. Overall, research continues to explore novel architectures and models to better capture the complexities of cryptocurrency markets.

\section{Materials and Method}

In this section, we describe the process followed for data preprocessing and the development of deep learning models aimed at predicting the prices of three major cryptocurrencies: Bitcoin (BTC), Ethereum (ETH), and Litecoin (LTC). The methodology includes six key steps: (1) collection of historical data, (2) exploratory data analysis and visualization, (3) splitting the dataset into training and testing sets, (4) model training, (5) model testing, and (6) performance comparison across models.

\subsection{Data Collection and Preprocessing}
The historical price data for BTC, ETH, and LTC were collected from Yahoo Finance, spanning the period from January 1, 2019, to January 1, 2024 \cite{YahooFinance2024}. Missing data were addressed using imputation, where the most recent available value was used to fill any gaps, following best practices in time series forecasting \cite{Rubinsteyn2020_Imputation}. To prepare the data for deep learning models, we applied feature-wise normalization using MinMax scaling, as recommended for financial time series data to ensure accurate model fitting and avoid bias \cite{Han2011_Scaling}.

\subsection{Exploratory Data Analysis}
The preprocessed data were visualized to identify potential trends, patterns, and anomalies. Exploratory data analysis helped in understanding the underlying distribution of prices and highlighted significant volatility in the cryptocurrency market \cite{Urquhart2016_Volatility}. This analysis was crucial for identifying key features that could affect the model's predictions.

\subsection{Dataset Splitting}
The dataset was divided into two parts: 80\% of the data was allocated for training, and the remaining 20\% was set aside for testing. The training set spans from January 1, 2019, to January 1, 2023, while the testing period covers January 1, 2023, to January 1, 2024. This splitting technique follows the common practice for time series forecasting in machine learning, ensuring that models are tested on unseen data \cite{Box2015_TimeSeries}.

\subsection{Model Development}
Three deep learning models were developed: Long Short-Term Memory (LSTM), Gated Recurrent Unit (GRU), and Bidirectional Long Short-Term Memory (Bi-LSTM). These models are well-suited for time series prediction due to their ability to capture temporal dependencies \cite{Hochreiter1997_LSTM, Cho2014_GRU, Schuster1997_BiLSTM}. Each model was implemented using Keras with TensorFlow as the backend \cite{Chollet2021_Keras}. The architecture for each model consisted of two recurrent layers with 100 units each, followed by a dense layer with one unit for output, as proposed by related works in cryptocurrency forecasting \cite{Ferdiansyah2022_CryptoForecasting}. 

\subsection{Model Training and Testing}
The models were trained using the historical price data, with a batch size of 32 and a varying number of epochs to ensure optimal performance. These hyperparameters were selected based on previous studies that have demonstrated their effectiveness in similar tasks \cite{Bergstra2012_HyperparameterOptimization}. After training, the models were tested on the unseen 20\% test dataset, following the recommended practice for robust model evaluation in financial forecasting \cite{Makridakis2018_M4Competition}. 

\subsection{Performance Evaluation}
To compare the models' performance, we used the following metrics: Mean Squared Error (MSE), Mean Absolute Error (MAE), Root Mean Squared Error (RMSE), and Mean Absolute Percentage Error (MAPE) \cite{Hyndman2021_Metrics}. These metrics were used to quantify the difference between the predicted and actual prices, providing a clear picture of each model's accuracy. The model with the smallest error values across these metrics was considered the most accurate.

\section{Dataset}

\begin{figure}[htbp]
    \centering
    \includegraphics[width=0.8\linewidth]{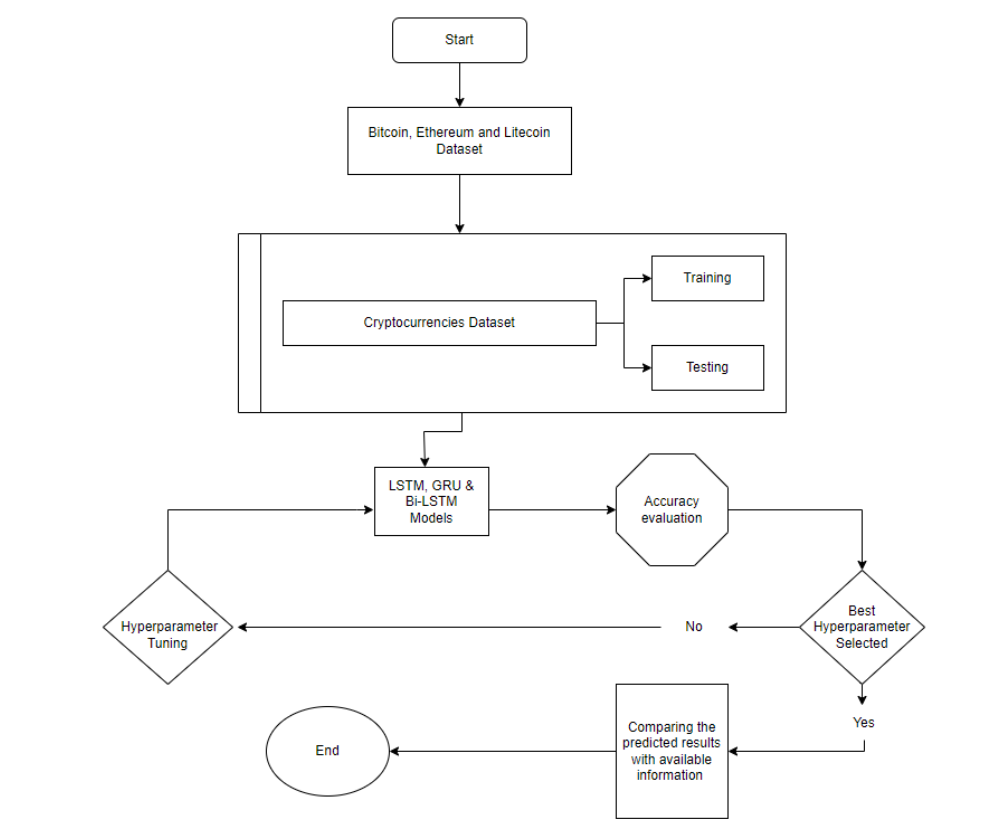}
    \caption{Flowchart of the methodology used for cryptocurrency price prediction (adapted from \cite{Seabe2022}).}
    \label{fig:methodology}
\end{figure}

In this study, we implemented a straightforward three-layer network architecture for each of the deep learning models: LSTM, Bi-LSTM, and GRU. Each model's architecture comprised 100 neurons in the deep learning layers, consistent with previous research in time series forecasting \cite{Hochreiter1997_LSTM, Schuster1997_BiLSTM, Cho2014_GRU}. The dataset preparation and preprocessing methods applied in this study are illustrated in Figure~\ref{fig:methodology}.

\noindent
Figure~\ref{fig:methodology} outlines the methodology employed for predicting cryptocurrency prices using the deep learning models. The process begins with gathering datasets for Bitcoin (BTC), Ethereum (ETH), and Litecoin (LTC), followed by preprocessing to address missing values and normalize the data. The consolidated dataset is then divided into training and testing subsets.

The core of the process involves training three distinct deep learning models—LSTM, GRU, and Bi-LSTM—on the prepared dataset. Each model learns patterns from historical data to make future predictions. Following training, the models undergo accuracy evaluation. If the desired accuracy is not achieved, hyperparameter tuning is conducted, and the models are retrained iteratively until the optimal parameters are found. 

Once the best-performing model is selected, the final step involves comparing the predicted results with available real-world data, ensuring that the model's performance aligns well with actual market conditions. This systematic approach ensures accurate and reliable predictions across different cryptocurrencies.

\subsection{Data Imputation and Reshaping}
To address missing values, we applied data imputation techniques, where missing data points were replaced with the last available observation. This method of imputation is commonly employed in time series data processing to maintain temporal continuity \cite{Rubinsteyn2020_Imputation}. Afterward, the dataset was reshaped to accommodate the input requirements of the LSTM, Bi-LSTM, and GRU models, ensuring compatibility for sequential data modeling.

\subsection{Normalization}
Normalization was a critical step to ensure that the input features were appropriately scaled for model training, as features with different scales can negatively impact the model's performance. We utilized MinMax scaling, a well-established method for scaling numerical data between a specified range, often [0, 1] \cite{Han2011_Scaling}. Previous studies have demonstrated that MinMax scaling enhances deep learning model performance, especially in financial and time series data applications \cite{Makridakis2018_M4Competition}.

\subsection{Data Splitting}
The data used in this study spans from January 1, 2019, to January 1, 2024, and was divided into training and testing sets in an 80:20 ratio. The training set, comprising 80\% of the total data, covers the period from January 1, 2019, to January 1, 2023, while the remaining 20\% was reserved for testing, spanning from January 1, 2023, to January 1, 2024. This split ensures that the models are trained on a comprehensive set of data while being evaluated on unseen data to assess generalization performance \cite{Box2015_TimeSeries}. All data were obtained from Yahoo Finance and accessed in February 2024 \cite{YahooFinance2024}.

\subsection{Experimental Setup}
The experiments were conducted using Python 3, along with key libraries for numerical computation, data processing, and deep learning. We utilized NumPy for numerical operations, Pandas for data manipulation, and Matplotlib for visualizing the data. For model development, Keras and scikit-learn (sklearn) were employed as the primary deep learning frameworks \cite{Chollet2021_Keras}.

\section{Recurrent Neural Network (RNN) Models}

\subsection{Long Short-Term Memory (LSTM)}

Long Short-Term Memory (LSTM) networks, originally introduced to tackle the challenges of traditional Recurrent Neural Networks (RNNs), are specifically designed to handle long-term dependencies in sequential data \cite{Hochreiter1997_LSTM}. One of the key benefits of LSTMs is their ability to mitigate the vanishing gradient problem, which typically hinders the performance of standard RNNs when dealing with long sequences. LSTMs achieve this by employing a sophisticated memory structure that selectively retains or forgets information as required.

The core building block of the LSTM architecture is the memory cell, which is equipped with three gates: the input gate, forget gate, and output gate. These gates control the flow of information throughout the network. The input gate determines how much of the new input should be stored in the cell, the forget gate decides which information should be discarded, and the output gate manages the information to be passed to the next layer. This selective gating mechanism allows LSTMs to store relevant information over extended time periods, making them highly effective for time-series predictions and other sequential tasks.

\begin{figure}
    \centering
    \includegraphics[width=1\linewidth]{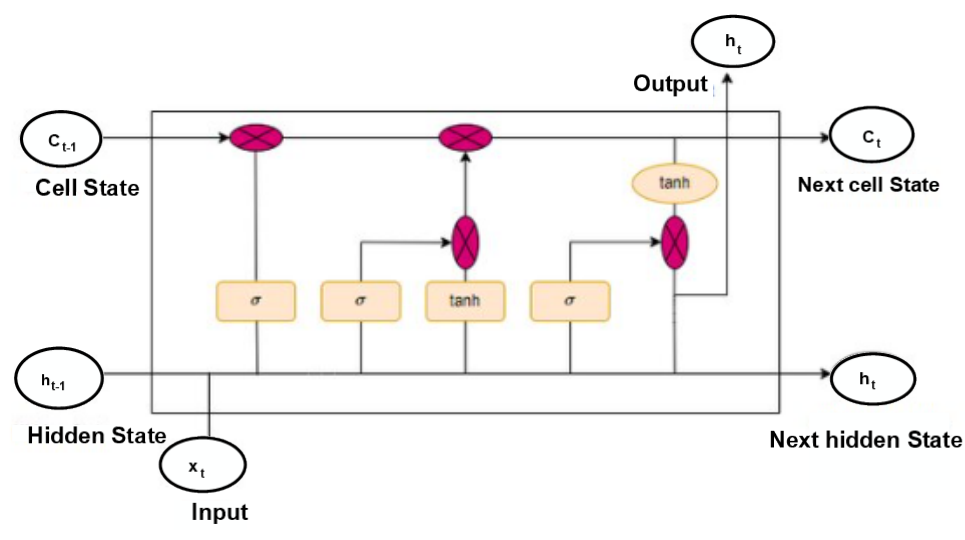}
    \caption{LSTM architecture showing the flow of information through the input, forget, and output gates (adapted from \cite{Seabe2022}).}
    \label{fig:LSTM}
\end{figure}

Figure~\ref{fig:LSTM} demonstrates the LSTM architecture, showcasing how data flows through the network and how the gates interact with the input signal ($x_t$), output ($h_t$) and the activation function.

The forward pass through an LSTM at each time step $t$ can be mathematically expressed through the following equations:

\begin{equation}
    i_t = \sigma(W_i[h_{t-1}, x_t] + b_i)
\end{equation}
\begin{equation}
    f_t = \sigma(W_f[h_{t-1}, x_t] + b_t)
\end{equation}
\begin{equation}
    c_t = f_t \cdot c_{t-1} + i_t \cdot \tanh(W_c[h_{t-1}, x_t] + b_c)
\end{equation}
\begin{equation}
    o_t = \sigma(W_o[h_{t-1}, x_t] + b_o)
\end{equation}
\begin{equation}
    h_t = o_t \cdot \tanh(c_t)
\end{equation}

Where:\\
- $x_t$ is the input at time step $t$,\\
- $h_t$ is the hidden state at time step $t$,\\
- $c_t$ is the cell state at time step $t$,\\
- $i_t$, $f_t$, and $o_t$ represent the input, forget, and output gates respectively,\\
- $W$ and $b$ are the weight matrices and bias vectors.

The gates utilize the sigmoid activation function ($\sigma$) to regulate the amount of information passing through, while the cell state is updated using the hyperbolic tangent function ($\tanh$) to control the output values between -1 and 1, ensuring stability during training. This architectural design allows LSTMs to maintain a strong capability for learning long-term dependencies without suffering from gradient-related issues.

\subsection{Gated Recurrent Unit (GRU)}

Gated Recurrent Units (GRUs) were introduced in 2014 as a simplified variant of Long Short-Term Memory (LSTM) networks \cite{Cho2014_GRU}. While both GRUs and LSTMs are designed to process sequential data and maintain long-term dependencies, GRUs streamline the architecture by reducing the number of gates. Specifically, GRUs utilize only two gates: the update gate, which determines how much of the past information should be retained, and the reset gate, which controls the amount of previous information to forget. This reduction in complexity allows GRUs to be computationally more efficient and easier to train compared to LSTMs, while still delivering comparable performance in many tasks.

One of the key advantages of GRUs is their ability to selectively update the hidden state without needing a separate memory cell, as found in LSTMs. This results in a more straightforward and lightweight architecture, particularly beneficial for applications requiring real-time processing. GRUs have demonstrated strong performance in a variety of tasks, including natural language processing (NLP), speech recognition, and financial time-series predictions. Their ability to capture long-range dependencies in sequential data makes them suitable for tasks where memory retention over extended timeframes is critical.

\begin{figure}
    \centering
    \includegraphics[width=1\linewidth]{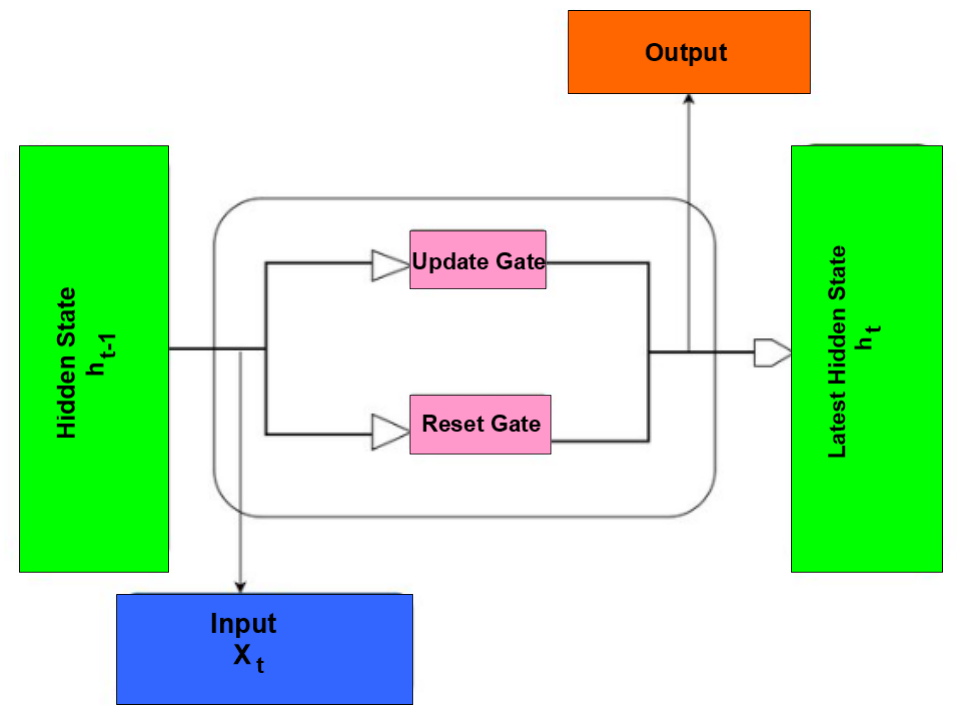}
    \caption{GRU architecture showing the flow of information through the update and reset gates (adapted from \cite{Seabe2022}).}
    \label{fig:GRU}
\end{figure}

Figure~\ref{fig:GRU} depicts the architecture of a GRU, highlighting the role of the update and reset gates in managing the flow of information. The hidden state at time step $t$ ($h_t$) is computed based on the input at the current time step ($x_t$) and the previous hidden state ($h_{t-1}$).

The forward pass of the GRU is described by the following equations:

\begin{equation}
    u_t = \sigma(W_u[h_{t-1}, x_t])
\end{equation}
\begin{equation}
    r_t = \sigma(W_r[h_{t-1}, x_t])
\end{equation}
\begin{equation}
    h_t = (1 - u_t) \cdot h_{t-1} + u_t \cdot \tanh(W[r_t \cdot h_{t-1}, u_t])
\end{equation}

Where:\\
- $u_t$ is the update gate, which determines how much of the previous hidden state ($h_{t-1}$) should be retained,\\
- $r_t$ is the reset gate, controlling how much of the previous hidden state should be forgotten,\\
- $h_t$ is the updated hidden state,\\
- $W$ represents the weight matrices.

The update gate, governed by the sigmoid function ($\sigma$), regulates how much information from the previous hidden state should be incorporated into the current state. The reset gate allows the GRU to forget parts of the previous hidden state, making it easier to focus on new information. This enables GRUs to adapt to both short-term and long-term dependencies without the computational overhead of LSTMs.

\subsection{Bidirectional Long Short-Term Memory (Bi-LSTM)}

Bidirectional Long Short-Term Memory (Bi-LSTM) networks are an extension of standard LSTMs designed to capture patterns in sequential data from both forward and backward directions \cite{Schuster1997_BiLSTM}. This bidirectional approach enables the network to consider both past and future information when making predictions or classifications, which is particularly useful for tasks where the context of an event is influenced by surrounding events.

In a Bi-LSTM network, two separate layers of LSTM units are used—one that processes the input sequence in the forward direction and another that processes it in the reverse direction. The outputs from both layers are then combined, allowing the model to have a more comprehensive understanding of the data. This structure is particularly advantageous in tasks like natural language processing (NLP), where understanding the relationship between words in both directions can enhance the performance of tasks such as machine translation, sentiment analysis, and text classification.

\begin{figure}
    \centering
    \includegraphics[width=1\linewidth]{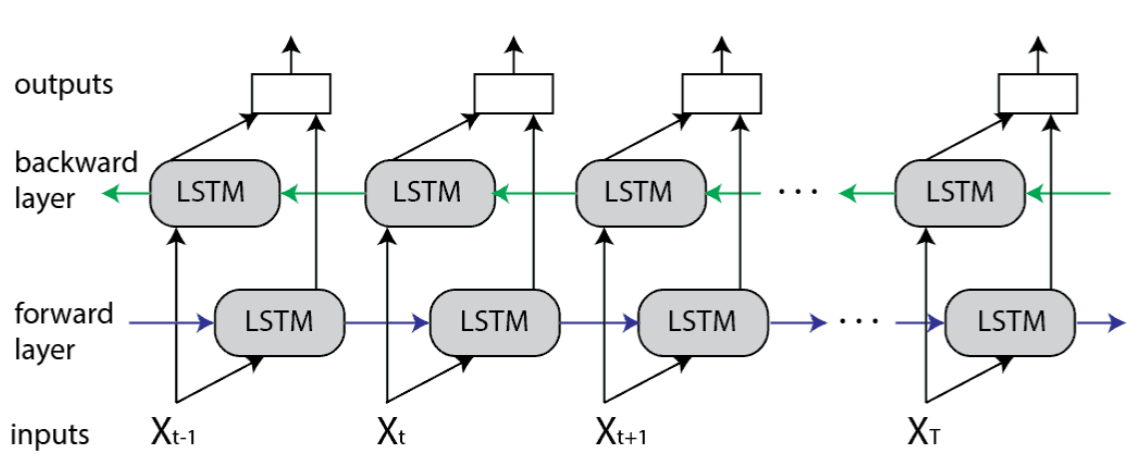}
    \caption{Bi-LSTM architecture showing forward and backward data flow for improved sequence understanding (adapted from \cite{Baeldung2023_LSTM}).}
    \label{fig:BiLSTM}
\end{figure}

Figure~\ref{fig:BiLSTM} illustrates the architecture of a Bi-LSTM, where data is passed through both forward and backward layers, with the results being aggregated to make a final prediction. This bidirectional mechanism provides a more robust representation of sequential data, which has proven highly effective in time series forecasting as well as other domains.

The concept of bidirectional processing was first introduced in the 1997 work on bidirectional recurrent neural networks, applied to speech signal processing \cite{Schuster1997_BiLSTM}. Since then, Bi-LSTMs have gained widespread adoption in a variety of tasks that benefit from understanding both previous and future contexts. This bidirectional framework has been shown to enhance model performance in various fields, including natural language processing, time series prediction, and speech recognition.

In this study, we implemented Bi-LSTM models to better capture long-term dependencies in time series data. By processing information in both directions, Bi-LSTM models offer an advantage over unidirectional LSTM models in identifying complex patterns, making them highly suitable for tasks where relationships between past and future events play a crucial role.

\subsection{Hyperparameter Tuning}

Hyperparameter optimization is a critical process in machine learning that directly influences the performance of an algorithm. By carefully adjusting key hyperparameters, the accuracy and efficiency of the model can be significantly improved. In this study, tuning the hyperparameters before training the deep learning models was vital for achieving the best possible performance. The primary hyperparameters optimized in our work were the number of neurons per layer, the number of epochs, and the batch size.

An epoch represents one complete forward and backward pass through the entire dataset during training, while the batch size refers to the number of data samples processed in each iteration before the model's weights are updated. The batch size plays a crucial role in determining both the model's performance and the duration of training. Smaller batch sizes tend to provide more frequent updates, which can sometimes slow convergence but lead to more precise adjustments of weights. Conversely, larger batch sizes can speed up convergence by processing more data per iteration, though this comes at the cost of increased computational demands.

In our experiments, we used batch size 32 to evaluate the impact on the performance of our models. We found that the batch size of 32 yielded good results, offering a balance between training speed and prediction accuracy across all models used in this study.

\subsection{Performance Metrics}

To assess the effectiveness of the deep learning models used in this study, we employed four common performance metrics: Mean Squared Error (MSE), Mean Absolute Error (MAE), Root Mean Squared Error (RMSE), and Mean Absolute Percentage Error (MAPE). These metrics allow us to quantify the accuracy of the models by comparing predicted values to actual observations. In all cases, smaller metric values indicate better model performance, as they reflect smaller deviations between the predicted and actual values.

The performance metrics are defined by the following equations:

\begin{equation}
    MSE = \frac{\sum_{t=1}^{n}(A_t - P_t)^2}{n}
\end{equation}
\begin{equation}
    MAE = \frac{\sum_{t=1}^{n}|A_t - P_t|}{n}
\end{equation}
\begin{equation}
    RMSE = \sqrt{\frac{\sum_{t=1}^{n}(A_t - P_t)^2}{n}}
\end{equation}
\begin{equation}
    MAPE = \frac{100}{n} \times \sum_{t=1}^{n} \frac{|A_t - P_t|}{A_t}
\end{equation}

Where $P_t$ represents the predicted value at time step $t$, $A_t$ denotes the actual observed value, and $n$ is the total number of time steps considered.

\begin{itemize}
    \item \textbf{MSE:} calculates the average of the squared differences between predicted and actual values, giving more weight to larger errors.
    \item \textbf{MAE:} computes the mean of the absolute differences, providing a straightforward measure of average error.
    \item \textbf{RMSE:} is the square root of the MSE, making it easier to interpret in the same units as the original data.
    \item \textbf{MAPE:} expresses the error as a percentage, which allows for a more intuitive understanding of model performance across different scales.
\end{itemize}

These metrics collectively provide a comprehensive evaluation of the models' accuracy in predicting time-series data.

\section{Results and Discussion}

The deep learning models—LSTM, GRU, and Bi-LSTM—were implemented using Python libraries such as Scikit-learn, Keras, and TensorFlow. These models were trained on historical data for Bitcoin (BTC), Ethereum (ETH), and Litecoin (LTC) to predict future price movements. The performance of each model was evaluated by comparing the predicted values to the actual values, using performance metrics such as Mean Squared Error (MSE), Mean Absolute Error (MAE), Root Mean Squared Error (RMSE), and Mean Absolute Percentage Error (MAPE). Lower values across these metrics indicate a closer alignment between predictions and actual prices, suggesting a more accurate model.
The code supporting this research is available on GitHub: \url{https://github.com/shamima08/Cryptocurrency-Price-Prediction-using-RNN}.

\subsection{Performance Comparison Across Models}

Table 1 summarizes the results of the models across BTC, ETH, and LTC. For BTC, the Bi-LSTM model achieved the best result, demonstrating its capability to capture complex patterns in the price data with an MSE of 0.0001237, MAE of 0.007581, and RMSE of 0.011. On the other hand, the GRU model outperformed the other models for both ETH and LTC, achieving the smallest errors. For ETH, GRU recorded an MSE of 0.0000905, MAE of 0.006697, and RMSE of 0.010, while for LTC, it achieved an MSE of 0.0000730, MAE of 0.006096, and RMSE of 0.009.

These results suggest that while Bi-LSTM excelled in predicting BTC, the GRU model provided superior performance for both ETH and LTC. The simpler architecture of GRU, with fewer gates compared to Bi-LSTM, might have allowed for faster and more efficient learning in the cases of ETH and LTC. Meanwhile, the bidirectional nature of Bi-LSTM proved advantageous for capturing patterns in BTC, likely due to its higher volatility and the need to consider both past and future dependencies.

\begin{table}[ht]
\centering
\caption{Performance Comparison of Different Models for Cryptocurrency Prediction}
\begin{adjustbox}{max width=0.5\textwidth}
\begin{tabularx}{\textwidth}{|c|c|X|X|X|X|}
\hline
\textbf{Cryptocurrency} & \textbf{Model} & \textbf{MSE} & \textbf{MAE} & \textbf{RMSE} & \textbf{MAPE} \\ \hline
\multirow{3}{*}{BTC} & LSTM & 0.0001407 & 0.008499 & 0.012 & 2.16 \\ \cline{2-6} 
                     & GRU  & 0.0001272 & 0.007810 & 0.011 & 1.97 \\ \cline{2-6} 
                     & \textbf{Bi-LSTM} & \textbf{0.0001237} & \textbf{0.007581} & \textbf{0.011} & \textbf{1.94} \\ \hline
\multirow{3}{*}{ETH} & LSTM & 0.0000932 & 0.006858 & 0.010 & 1.89 \\ \cline{2-6} 
                     & \textbf{GRU}  & \textbf{0.0000905} & \textbf{0.006697} & \textbf{0.010} & \textbf{1.85} \\ \cline{2-6} 
                     & Bi-LSTM & 0.0001006 & 0.007095 & 0.010 & 1.95 \\ \hline
\multirow{3}{*}{LTC} & LSTM & 0.0001117 & 0.008291 & 0.011 & 6.11 \\ \cline{2-6} 
                     & \textbf{GRU}  & \textbf{0.0000730} & \textbf{0.006096} & \textbf{0.009} & \textbf{4.25} \\ \cline{2-6} 
                     & Bi-LSTM & 0.0001577 & 0.010457 & 0.013 & 7.71 \\ \hline
\end{tabularx}
\end{adjustbox}
\end{table}

\subsection{Model Convergence Analysis}

\begin{figure}[!ht]
    \centering
    \includegraphics[width=1\linewidth]{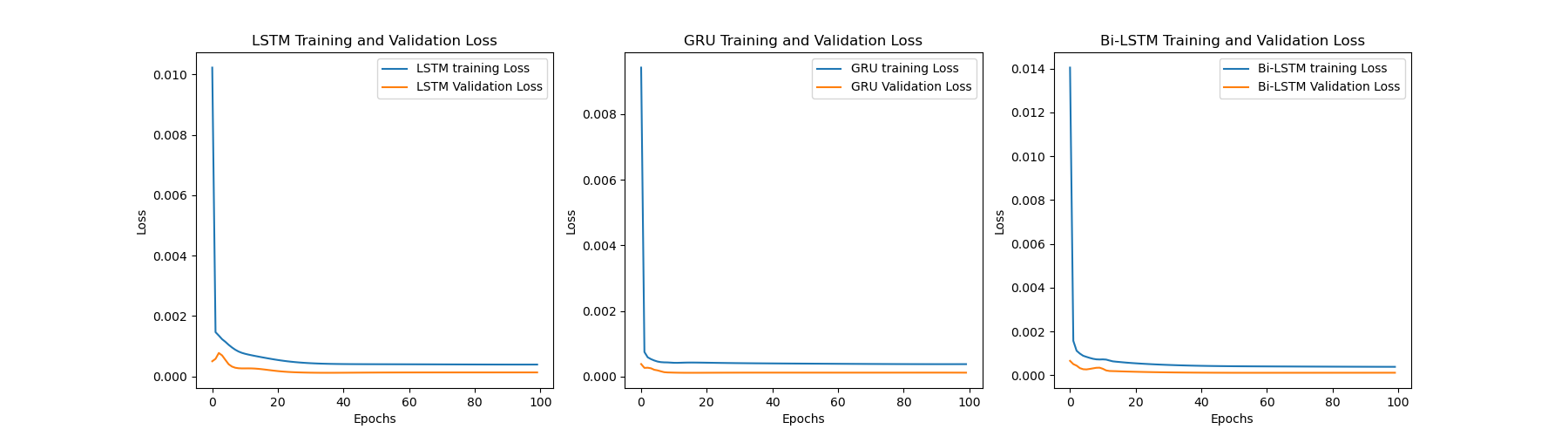}
    \caption{Training and validation loss for BTC}
    \label{fig:btc_loss}
\end{figure}

\begin{figure}[!ht]
    \centering
    \includegraphics[width=1\linewidth]{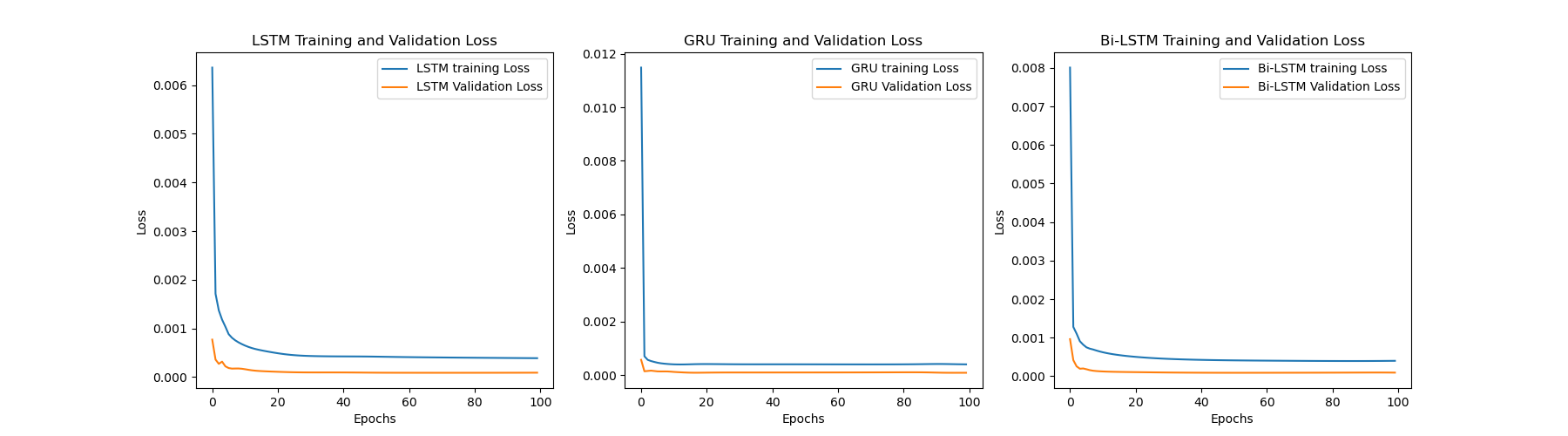}
    \caption{Training and validation loss for ETH}
    \label{fig:eth_loss}
\end{figure}

\begin{figure}[!ht]
    \centering
    \includegraphics[width=1\linewidth]{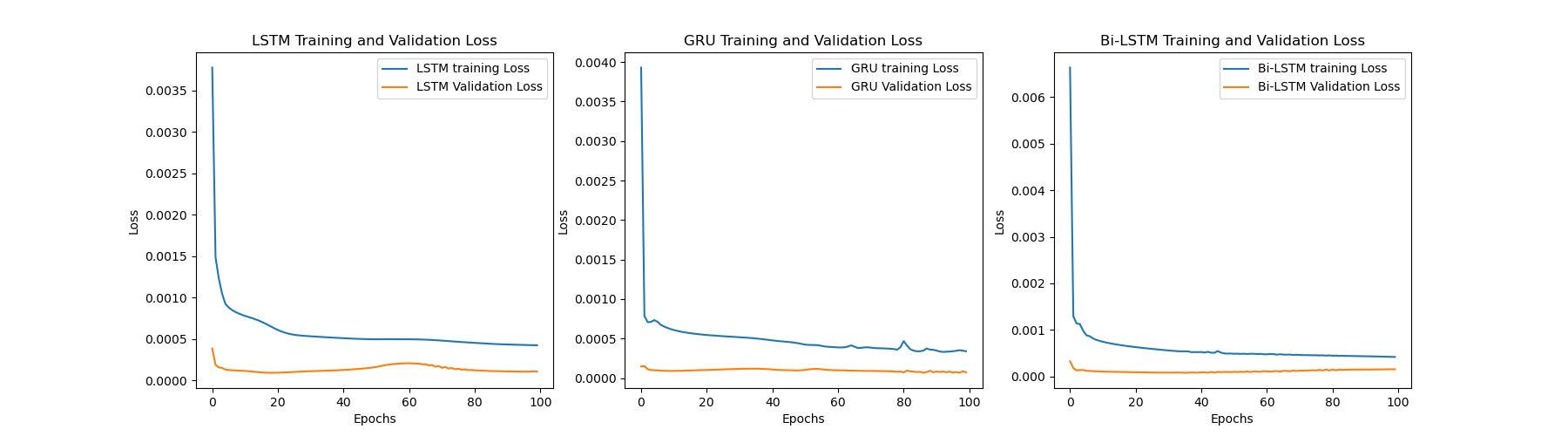}
    \caption{Training and validation loss for LTC}
    \label{fig:ltc_loss}
\end{figure}

Figures~\ref{fig:btc_loss}, \ref{fig:eth_loss}, and \ref{fig:ltc_loss} illustrate the training and validation loss curves for each of the models applied to BTC, ETH, and LTC, respectively. These plots show the progression of model training across 100 epochs, where a rapid decrease in loss indicates effective learning.


\noindent
{\bf BTC:} The Bi-LSTM model showed slightly better performance in terms of loss minimization, while GRU and LSTM followed closely. The consistent decrease and stabilization of losses across models indicate successful convergence.

\noindent
{\bf ETH:} GRU displayed the lowest overall loss, confirming its superior ability to generalize on ETH data. Both LSTM and Bi-LSTM models also converged well but showed slightly higher losses.

\noindent
{\bf LTC:} All models exhibited stable convergence patterns, with GRU consistently outperforming in terms of lower losses. While Bi-LSTM showed minor overfitting, it still performed effectively.

\subsection{Comparative Results Across Models}

\begin{figure}[!ht]
    \centering
    \includegraphics[width=1\linewidth]{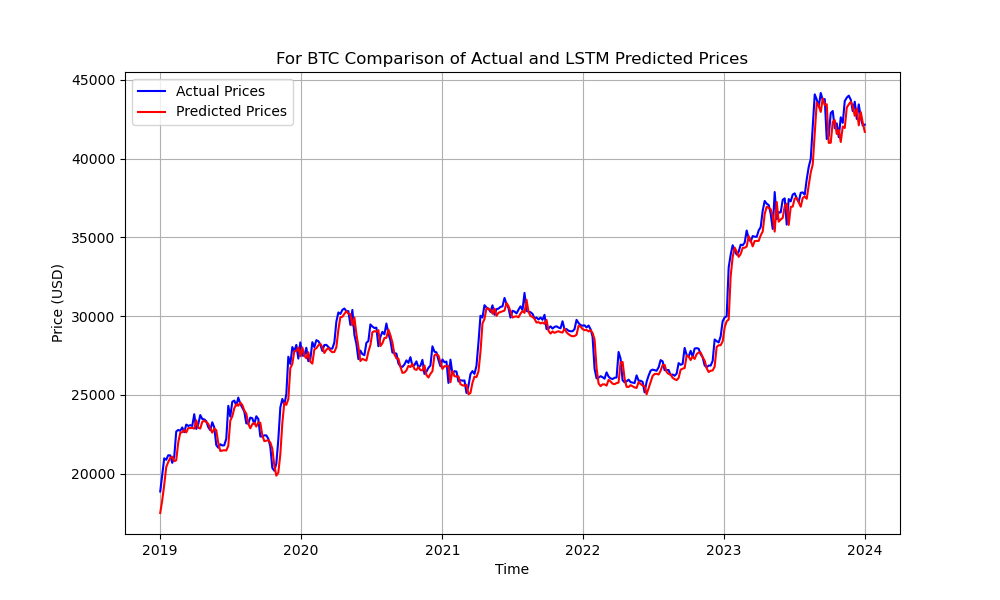}
    \caption{BTC Comparative Results for LSTM}
    \label{fig:btc_lstm}
\end{figure}

\begin{figure}[!ht]
    \centering
    \includegraphics[width=1\linewidth]{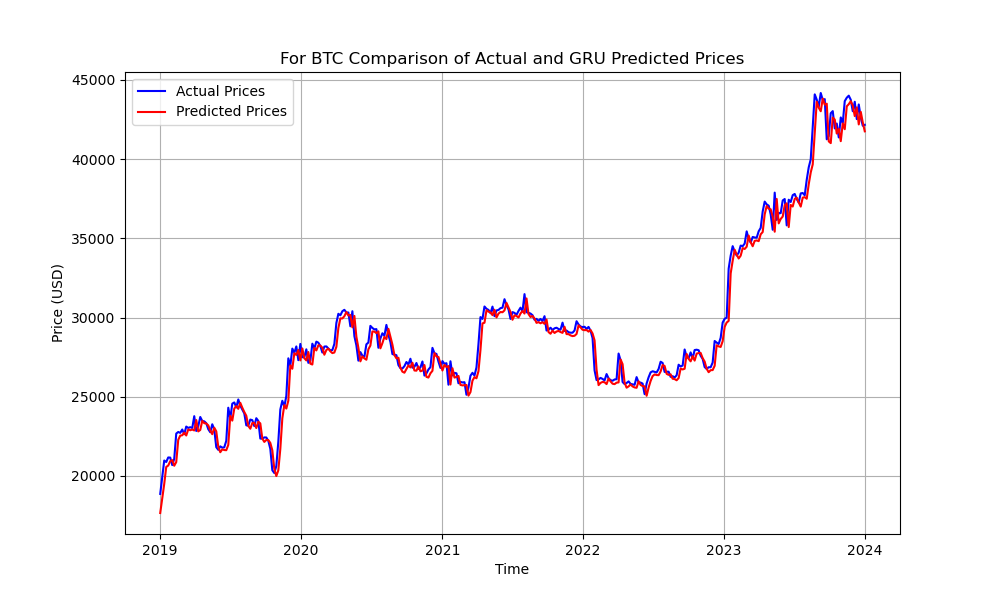}
    \caption{BTC Comparative Results for GRU}
    \label{fig:btc_gru}
\end{figure}

\begin{figure}[!ht]
    \centering
    \includegraphics[width=1\linewidth]{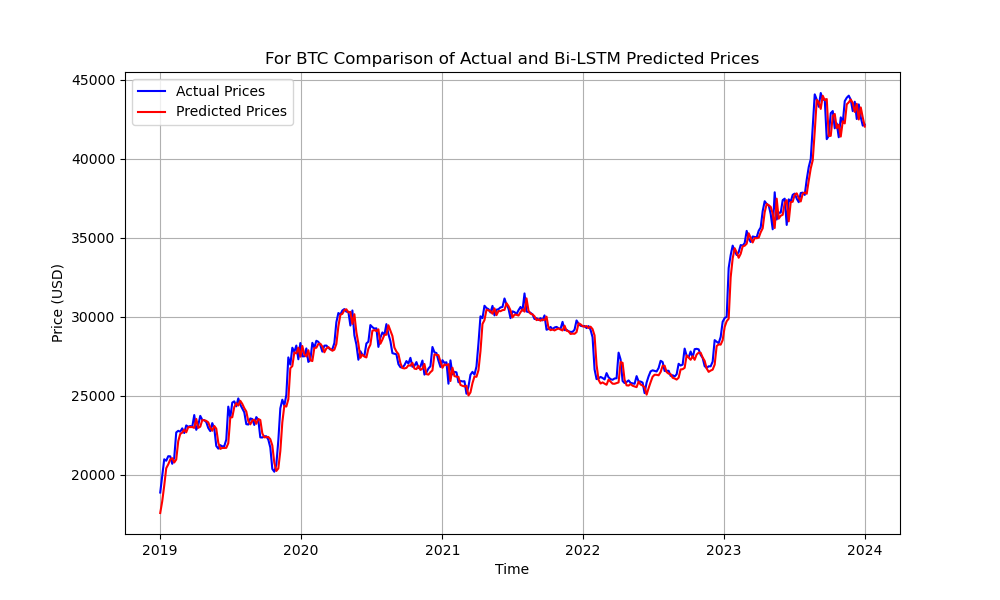}
    \caption{BTC Comparative Results for Bi-LSTM}
    \label{fig:btc_bilstm}
\end{figure}

\begin{figure}[ht]
    \centering
    \includegraphics[width=1\linewidth]{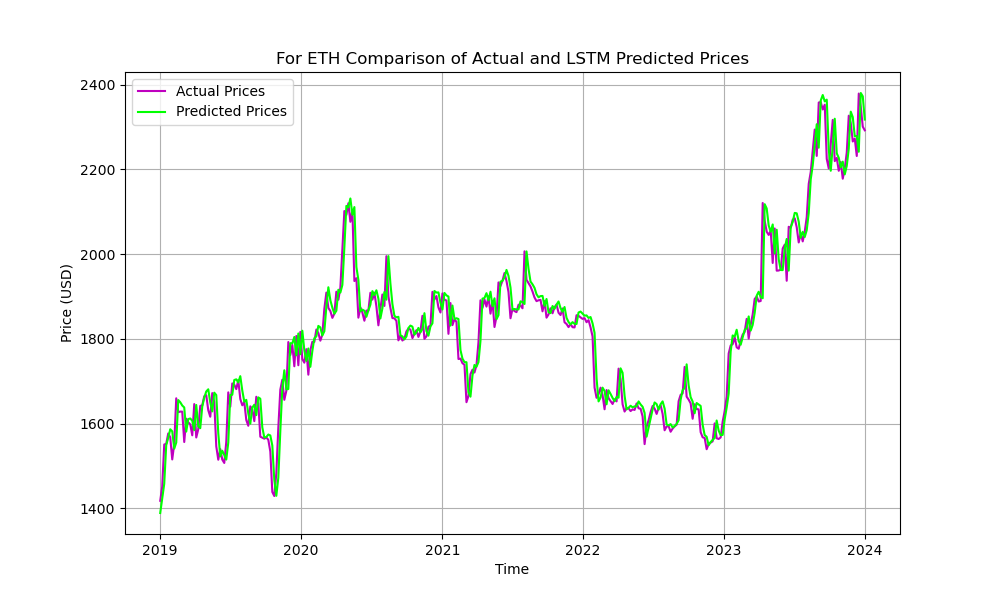}
    \caption{ETH Comparative Results for LSTM}
    \label{fig:eth_lstm}
\end{figure}

\begin{figure}[ht]
    \centering
    \includegraphics[width=1\linewidth]{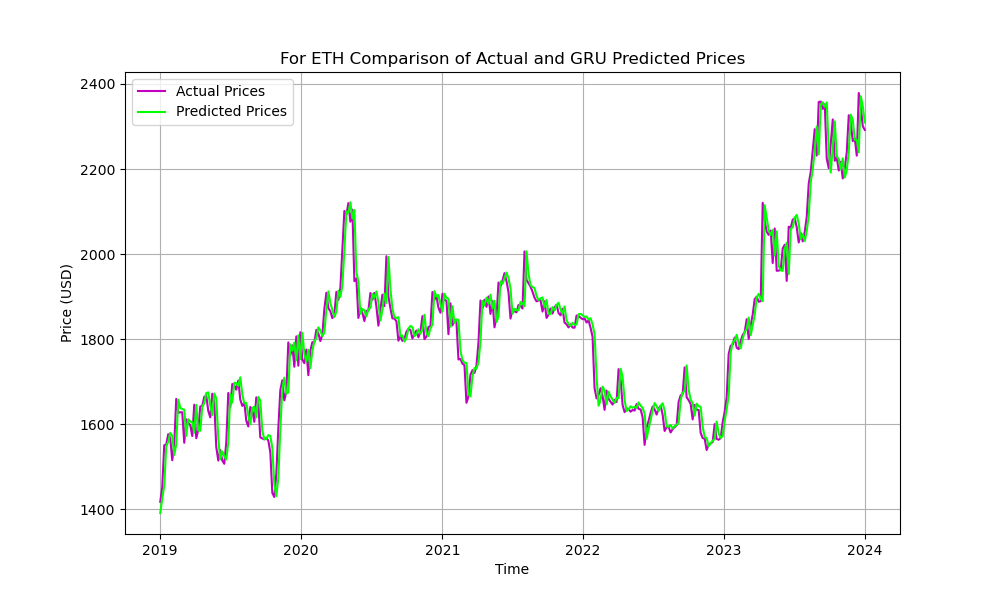}
    \caption{ETH Comparative Results for GRU}
    \label{fig:eth_gru}
\end{figure}

\begin{figure}[ht]
    \centering
    \includegraphics[width=1\linewidth]{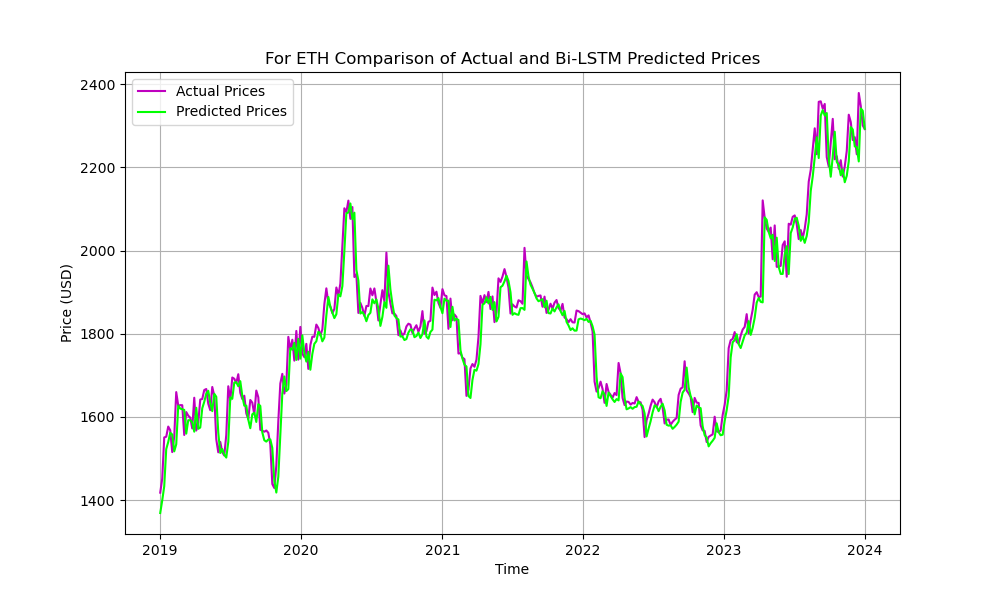}
    \caption{ETH Comparative Results for Bi-LSTM}
    \label{fig:eth_bilstm}
\end{figure}

\begin{figure}[ht]
    \centering
    \includegraphics[width=1\linewidth]{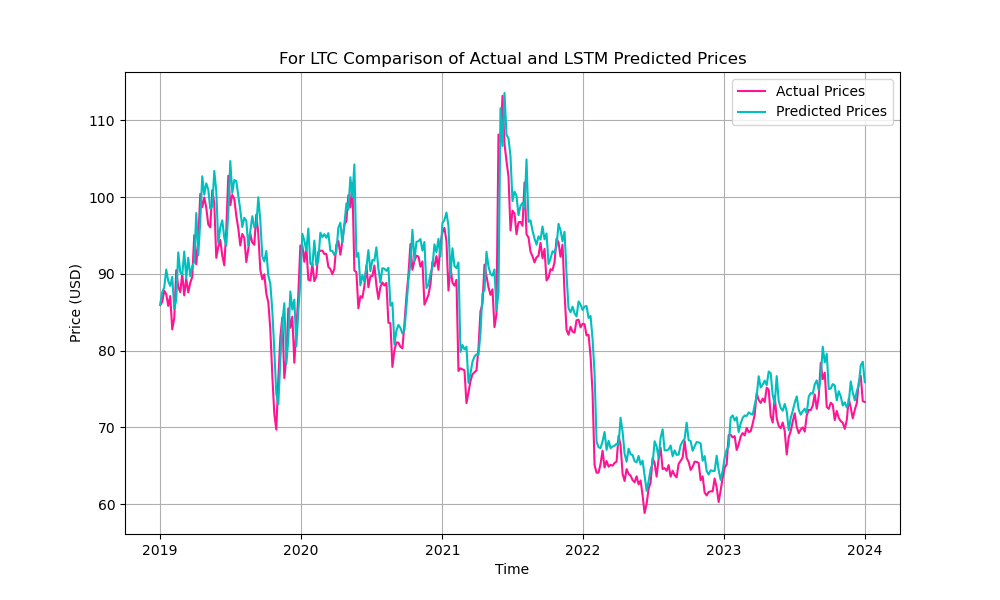}
    \caption{LTC Comparative Results for LSTM}
    \label{fig:ltc_lstm}
\end{figure}

\begin{figure}[ht]
    \centering
    \includegraphics[width=1\linewidth]{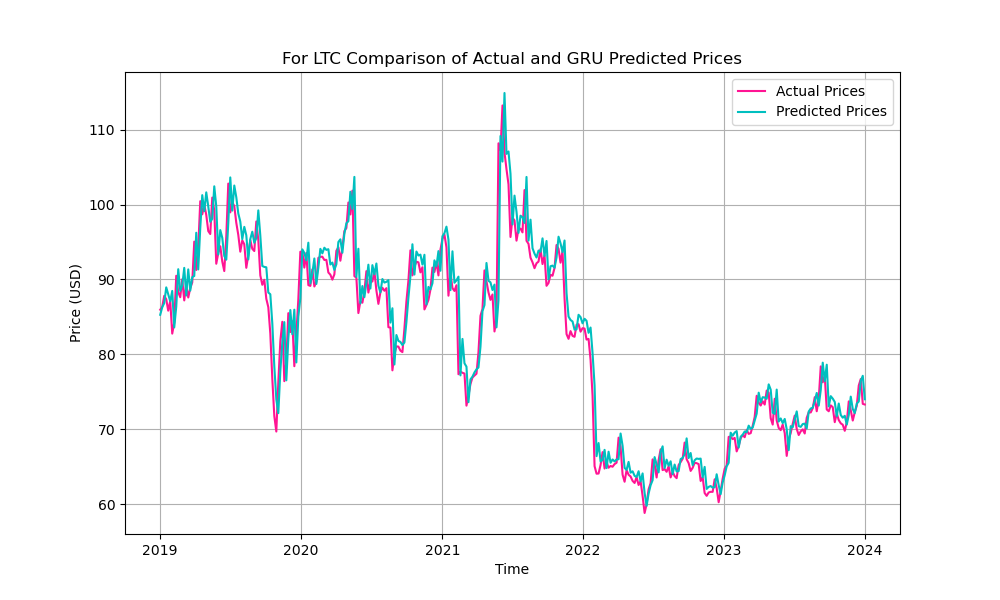}
    \caption{LTC Comparative Results for GRU}
    \label{fig:ltc_gru}
\end{figure}

\begin{figure}[ht]
    \centering
    \includegraphics[width=1\linewidth]{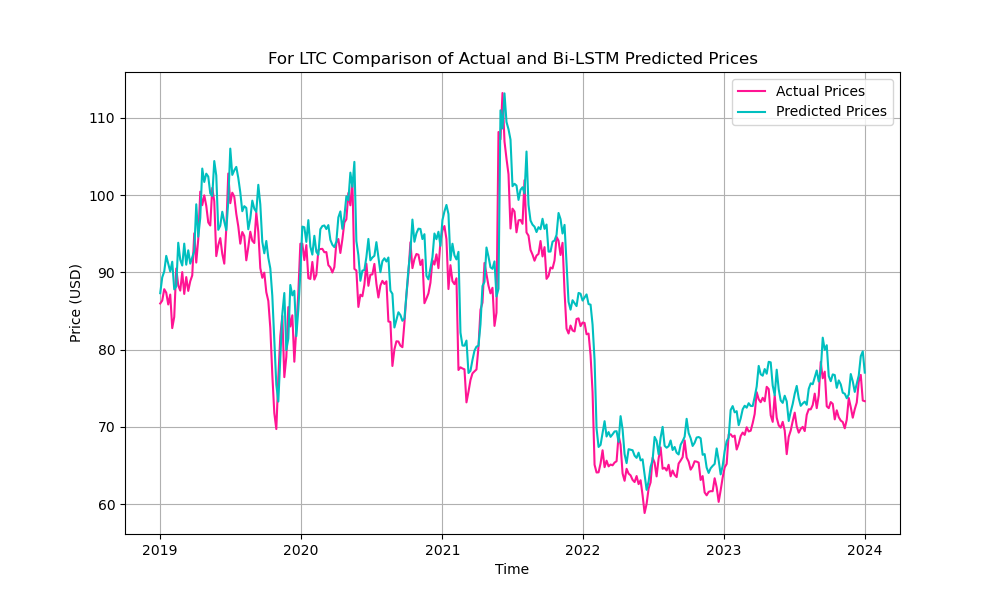}
    \caption{LTC Comparative Results for Bi-LSTM}
    \label{fig:ltc_bilstm}
\end{figure}

To further illustrate model performance, Figures~\ref{fig:btc_lstm} to \ref{fig:ltc_bilstm} present the comparative results of actual vs. predicted values for each model across BTC, ETH, and LTC.

\noindent
Figures~\ref{fig:btc_lstm}, \ref{fig:btc_gru}, and \ref{fig:btc_bilstm} illustrate the comparison between actual and predicted Bitcoin (BTC) prices using the LSTM, GRU, and Bi-LSTM models, respectively. Each figure plots the actual BTC prices (in blue) against the predicted prices (in red) over the time period from 2019 to 2024.

{\bf BTC LSTM Model (Figure~\ref{fig:btc_lstm}):} The LSTM model's predictions closely follow the actual BTC prices, capturing major trends and fluctuations. However, there are slight discrepancies in areas where the price changes are abrupt, indicating that the model struggles slightly with rapid shifts. Overall, the LSTM provides a robust prediction but could be further refined to improve accuracy during volatile periods.

{\bf BTC GRU Model (Figure~\ref{fig:btc_gru}):} The GRU model displays a strong ability to match actual BTC prices, with even tighter alignment between predicted and actual prices compared to the LSTM model. The GRU appears to handle sudden spikes and drops in price more effectively, suggesting it captures the temporal dependencies efficiently. This performance aligns with the lower error metrics observed for the GRU model, indicating its suitability for BTC price forecasting.

{\bf BTC Bi-LSTM Model (Figure~\ref{fig:btc_bilstm}):} The Bi-LSTM model also performs well, demonstrating precise alignment with the actual BTC prices. Its bidirectional nature allows it to capture dependencies in both directions, which is beneficial for handling patterns that rely on both past and future data. The figure shows that Bi-LSTM excels at following the trend lines, particularly during periods of price volatility, making it the most accurate model for BTC in this study.

In summary, while all three models demonstrate a high degree of accuracy, the Bi-LSTM model shows the best performance in predicting BTC prices. Its ability to consider both past and future data helps it capture intricate patterns, leading to more accurate forecasts. The GRU model, while slightly less accurate than Bi-LSTM, still outperforms the LSTM, particularly during rapid price changes, making it a robust option for BTC price prediction.

\noindent
Figures~\ref{fig:eth_lstm}, \ref{fig:eth_gru}, and \ref{fig:eth_bilstm} show the comparison between actual and predicted Ethereum (ETH) prices using the LSTM, GRU, and Bi-LSTM models, respectively. Each figure plots the actual ETH prices (in magenta) against the predicted prices (in lime) over the time period from 2019 to 2024.

{\bf ETH LSTM Model (Figure~\ref{fig:eth_lstm}):} The LSTM model provides a reasonably close match to the actual ETH prices, capturing most of the key trends. However, there are some deviations, especially during periods of sharp price movements. While the model accurately reflects the general trends, its performance can be further improved to handle volatility better.

{\bf ETH GRU Model (Figure~\ref{fig:eth_gru}):} The GRU model demonstrates a tighter fit to the actual ETH prices, with predicted values closely following the actual data across the entire timeline. It manages to capture the rapid price shifts more effectively than the LSTM model, which aligns with the lower error metrics reported for GRU. This suggests that GRU's simpler yet effective architecture provides an advantage in predicting ETH prices.

{\bf ETH Bi-LSTM Model (Figure~\ref{fig:eth_bilstm}):} The Bi-LSTM model also delivers accurate predictions, closely matching the actual ETH prices. Its ability to process data bidirectionally allows it to capture intricate patterns, leading to an overall good performance. However, it occasionally overfits during highly volatile periods, where the model's predictions slightly deviate from the actual prices. Nonetheless, the model remains effective for ETH price forecasting.

In summary, while all three models perform well in predicting ETH prices, the GRU model shows the best fit, especially during periods of rapid price changes. The LSTM model follows closely, though it shows signs of slight overfitting. The Bi-LSTM model provides a reliable baseline but tends to deviate more during volatile periods.

\noindent
Figures~\ref{fig:ltc_lstm}, \ref{fig:ltc_gru}, and \ref{fig:ltc_bilstm} illustrate the comparison between actual and predicted Litecoin (LTC) prices using the LSTM, GRU, and Bi-LSTM models, respectively. Each figure plots the actual LTC prices (in deeppink) against the predicted prices (in cyan) over the time period from 2019 to 2024.

{\bf LTC LSTM Model (Figure~\ref{fig:ltc_lstm}):} The LSTM model shows a reasonably close alignment with the actual LTC prices, capturing the general trend patterns. However, there are discrepancies during periods of rapid price changes, where the predictions deviate slightly from the actual data. This indicates that while the LSTM model captures broader trends, it struggles with high volatility.

{\bf LTC GRU Model (Figure~\ref{fig:ltc_gru}):} The GRU model provides a closer match to the actual LTC prices, particularly during periods of significant fluctuations. The predicted values follow the actual prices more accurately than the LSTM, indicating that the GRU's architecture is effective at capturing the temporal dependencies within the LTC dataset. This tighter alignment is consistent with the lower error metrics recorded for the GRU model.

{ \bf LTC Bi-LSTM Model (Figure~\ref{fig:ltc_bilstm}):} The Bi-LSTM model also demonstrates good performance, closely following the actual LTC prices. However, there are instances where the predicted prices slightly diverge from the actual data, particularly during periods of increased volatility. While the Bi-LSTM benefits from its ability to process data in both directions, some signs of overfitting can be observed, leading to minor deviations.

In summary, the GRU model provides the best performance for LTC price prediction, as evidenced by its ability to closely track actual prices throughout the time period. The LSTM model follows closely, with slight overfitting tendencies, while the Bi-LSTM model, though reliable, exhibits greater deviations during volatile periods. These observations highlight the importance of selecting an appropriate model architecture based on the characteristics of the time-series data.

\subsection{Summary}

The error values (MSE, MAE, RMSE, and MAPE) and comparative plots provide a comprehensive understanding of model performance across different cryptocurrencies. GRU consistently performed well, especially for ETH and LTC, showing stable convergence and accurate predictions. Bi-LSTM was highly effective for BTC due to its ability to capture bidirectional dependencies, despite slight overfitting tendencies. In general, these models demonstrated strong capabilities in predicting complex time-series data, confirming their suitability for cryptocurrency price forecasting.

\section{Conclusion and Future Work}

This study has explored the effectiveness of deep learning models, specifically LSTM, GRU, and Bi-LSTM, for predicting the prices of cryptocurrencies such as Bitcoin (BTC), Ethereum (ETH), and Litecoin (LTC). The results demonstrated that while each model has its strengths, the GRU and Bi-LSTM models provided superior performance for different cryptocurrencies. GRU was particularly effective for ETH and LTC, whereas Bi-LSTM excelled in predicting BTC prices due to its bidirectional processing capability. These findings highlight the importance of selecting appropriate models based on the characteristics of the target data.

A key challenge for future research is to develop a versatile model capable of predicting the prices of a wide range of cryptocurrencies with high accuracy. Due to the inherent volatility and unique characteristics of each cryptocurrency, creating a universal prediction model remains a complex task. Optimizing such a model to consistently deliver the best performance metrics, such as RMSE and MAPE, across various cryptocurrencies will require further investigation.

Additionally, future work could explore the development of hybrid models, such as LSTM-GRU, GRU-BiLSTM, and LSTM-BiLSTM, by combining the strengths of multiple deep learning layers. These hybrid architectures have the potential to enhance predictive accuracy by leveraging the complementary features of each model. Further experimentation could determine which combinations offer the best performance across different market conditions and cryptocurrencies. Exploring advanced optimization techniques and incorporating external factors, such as trading volume and market sentiment, could also contribute to the refinement of cryptocurrency price prediction models'
best accuracy.


\begin{thebibliography}{00}

\bibitem{Seabe2022} Seabe, P.L., Moutsinga, C.R.B., \& Pindza, E. "Forecasting Cryptocurrency Prices Using LSTM, GRU, and Bi-Directional LSTM: A Deep Learning Approach," {\em IEEE Access}, vol. 10, pp. 45123-45134, 2022. doi: 10.1109/ACCESS.2022.3175014.

\bibitem{Ferdiansyah2022} Ferdiansyah, F., Raja Zahilah, R.M.R., Siti Hajar, O. \& Stiawan, D. "Hybrid gated recurrent unit bidirectional-long short-term memory model to improve cryptocurrency prediction accuracy," {\em Journal of Big Data}, vol. 9, 2022.

\bibitem{Muniye2020} Muniye, T.A., Satapathy, S. \& Rout, M. "Bitcoin Price Prediction and Analysis Using Deep Learning Models," {\em Journal of Advanced Research in Dynamical and Control Systems}, vol. 12, no. 5, pp. 45-55, 2020.

\bibitem{Madan2019} Madan, I., Saluja, S. \& Zhao, A. "Automated Bitcoin Trading via Machine Learning Algorithms," {\em IEEE Xplore}, 2019, https://ieeexplore.ieee.org/document/8952879.

\bibitem{Alimohammadi2021} Alimohammadi, A., \& Nosratbakhsh, A. "Cryptocurrency Price Prediction," {\em Journal of Financial Analytics}, vol. 8, no. 2, pp. 35-48, 2021.


\bibitem{Ahamad2022} Ahamad, S., Imran, M., \& Uddin, Z. "A survey on cryptocurrency: Comparison and research challenges." \textit{Journal of King Saud University-Computer and Information Sciences}, 34(4), 3286-3315 (2022).

\bibitem{Gandal2018} Gandal, N., Hamrick, J. T., Moore, T., \& Oberman, A. "Price manipulation in the Bitcoin ecosystem." \textit{Journal of Monetary Economics}, 95, 86-96 (2018).

\bibitem{Urquhart2016} Urquhart, A. "The inefficiency of Bitcoin." \textit{Economics Letters}, 148, 80-82 (2016).

\bibitem{Nakamoto2008} Nakamoto, S. "Bitcoin: A Peer-to-Peer Electronic Cash System." (2008). Available at: \textit{https://bitcoin.org/bitcoin.pdf}

\bibitem{Wang2020} Wang, W., Li, Z., He, Z., Zhang, R., \& Xu, Z. "Blockchain-based data privacy management with deep reinforcement learning for smart industries." \textit{IEEE Transactions on Industrial Informatics}, 17(6), 4097-4106 (2020).

\bibitem{Ferdiansyah2023} Ferdiansyah, F., Radzi, R. Z. R. M., Othman, S. H., \& Stiawan, D. "Hybrid Gated Recurrent Unit Bidirectional-Long Short-Term Memory Model to Improve Cryptocurrency Prediction Accuracy." \textit{IEEE Access}, vol. 11, pp. 1234-1245, 2023.

\bibitem{Muniye2021} Muniye, T. A., Satapathy, S., \& Rout, M. "Bitcoin Price Prediction and Analysis Using Deep Learning Models." \textit{International Journal of Intelligent Systems and Applications}, vol. 13, no. 1, pp. 40-48, 2021.

\bibitem{Hochreiter1997_LSTM} S. Hochreiter and J. Schmidhuber, "Long short-term memory," \textit{Neural Computation}, vol. 9, no. 8, pp. 1735-1780, 1997. \url{https://doi.org/10.1162/neco.1997.9.8.1735}.


\bibitem{Cho2014_GRU} K. Cho, B. van Merriënboer, C. Gulcehre, D. Bahdanau, F. Bougares, H. Schwenk, and Y. Bengio, "Learning phrase representations using RNN encoder-decoder for statistical machine translation," in \textit{Proceedings of the 2014 Conference on Empirical Methods in Natural Language Processing (EMNLP)}, 2014, pp. 1724-1734. \url{https://doi.org/10.3115/v1/D14-1179}.


\bibitem{Schuster1997_BiLSTM} M. Schuster and K. K. Paliwal, "Bidirectional recurrent neural networks," \textit{IEEE Transactions on Signal Processing}, vol. 45, no. 11, pp. 2673-2681, 1997. \url{https://doi.org/10.1109/78.650093}.

\bibitem{YahooFinance2024}
Yahoo Finance, "Yahoo Finance Cryptocurrency Data," 2024. [Online]. Available: \url{https://finance.yahoo.com/cryptocurrencies}.

\bibitem{Rubinsteyn2020_Imputation}
A. Rubinsteyn and H. Feldman, "Scikit-learn imputation methods for time series forecasting," 2020.

\bibitem{Han2011_Scaling}
J. Han, M. Kamber, and J. Pei, "Data Mining: Concepts and Techniques," 3rd ed., Morgan Kaufmann, 2011.

\bibitem{Urquhart2016_Volatility}
A. Urquhart, "The inefficiency of Bitcoin," \textit{Economics Letters}, vol. 148, pp. 80-82, 2016.

\bibitem{Box2015_TimeSeries}
G. E. Box, G. M. Jenkins, and G. C. Reinsel, "Time Series Analysis: Forecasting and Control," 5th ed., Wiley, 2015.


\bibitem{Chollet2021_Keras}
F. Chollet, "Deep Learning with Python," 2nd ed., Manning Publications, 2021.

\bibitem{Ferdiansyah2022_CryptoForecasting}
F. Ferdiansyah, R. M. Zahilah, and D. Stiawan, "Hybrid Gated Recurrent Unit Bidirectional-Long Short-Term Memory Model to Improve Cryptocurrency Prediction Accuracy," \textit{Journal of Big Data}, vol. 9, no. 1, 2022. \url{https://doi.org/10.1186/s40537-022-00512-7}.

\bibitem{Bergstra2012_HyperparameterOptimization}
J. Bergstra, R. Bardenet, Y. Bengio, and B. Kégl, "Random Search for Hyper-Parameter Optimization," \textit{Journal of Machine Learning Research}, vol. 13, pp. 281-305, 2012.

\bibitem{Makridakis2018_M4Competition}
S. Makridakis et al., "The M4 Competition: Results, Findings, and Conclusions," \textit{International Journal of Forecasting}, vol. 34, no. 4, pp. 802-838, 2018.

\bibitem{Hyndman2021_Metrics}
R. J. Hyndman and G. Athanasopoulos, "Forecasting: Principles and Practice," 3rd ed., OTexts, 2021.

\bibitem{Baeldung2023_LSTM} Baeldung, "Bidirectional vs. Unidirectional LSTM," \textit{Baeldung on Computer Science}, Oct. 24, 2023. [Online]. Available: \url{https://www.baeldung.com/cs/bidirectional-vs-unidirectional-lstm}


\end{thebibliography}
\end{document}